\documentclass[12pt,a4paper,oneside]{article}
\usepackage{mathrsfs}
\usepackage[hmargin=2.3cm,vmargin=2.3cm]{geometry}
 \usepackage{graphicx}
\usepackage{amsmath,amssymb,amsthm,mathtools}
\usepackage{amsthm}
\newtheorem{theorem}{Theorem}
\newtheorem{corollary}[theorem]{Corollary}
\usepackage{bm}
  \usepackage{enumerate}
\usepackage{cite}
\usepackage{mathabx}
\usepackage{url}
\usepackage{hhline}
\newcommand{\bey}{\begin{eqnarray}}
\newcommand{\eey}{\end{eqnarray}}
\usepackage{etoolbox}

\usepackage{accents}
\usepackage{bm}

\theoremstyle{definition}

\usepackage{lipsum}

\let\OLDthebibliography\thebibliography
\renewcommand\thebibliography[1]{
  \OLDthebibliography{#1}
  \setlength{\parskip}{0pt}
  \setlength{\itemsep}{2pt plus 0.3ex}
}
 \usepackage{enumitem}
\setlist[itemize]{leftmargin=1.2em}
\setlist[enumerate]{leftmargin=1.2em}

\begin{document}

\title{Detection times are not  local observables in quantum field theory }
\author { Charis Anastopoulos\footnote{anastop@upatras.gr} \;   and   \; Ntina Savvidou\footnote{ksavvidou@upatras.gr}\\
 {\small Laboratory of Universe Sciences, Department of Physics, University of Patras, 26500 Greece} }
\maketitle

\begin{abstract}
Detection times are routinely recorded in quantum experiments.
We show that, under natural covariance and consistency
requirements, the observables describing these times cannot
be algebraically localized in any bounded spacetime region,
even when the records themselves are localized. This challenges
the identification of algebraic localization with the physical
localization of measurements and shows why microcausality
alone does not establish non-signalling for  
measurements in quantum field theory.
\end{abstract}

\vspace{0.6cm}

Time of detection is a ubiquitous observable in quantum experiments. It can be the principal recorded quantity, as in measurements of arrival times, decay times and time-resolved fluorescence \cite{Hirv, ICARUS}, or enter temporal correlations, such as those revealing photon bunching and antibunching \cite{HBT, QOptics}. More generally, experiments that resolve individual detection events routinely record timestamps \cite{Michalet, LHCb}, making detection time part of the measurement outcome.

 Here we show that, in relativistic quantum field theory (QFT), the quantum observable describing when a detection occurs cannot be confined to the spacetime region in which it is recorded—or to any finite enlargement of that region. This raises the question of whether the conventional description of local observables in QFT suffices to describe complete measurement records that include the times at which they are produced. 

Time observables generally require a description in terms of
positive operator-valued measures (POVMs). This is because a
Hamiltonian bounded from below cannot admit a self-adjoint time
operator that is covariant under time
translations. POVMs avoid this restriction and have been
constructed for time observables in  
many different contexts---see, for example, Refs. \cite{Kijow, Werner, Freden, QTP1}.

We consider detection times after a Cauchy surface, labelled $t = 0$, by which quantum state 
preparation has been completed. For a finite recording duration $T$, a detection-time POVM
assigns a positive operator $\hat{\Pi}_T(\Delta)$ to each
measurable set $\Delta\subset(0,T]$. In a state $\hat{\rho}$,
the probability of detection during $\Delta$ is
\begin{equation}
P_T(\Delta)
=\operatorname{Tr}\!\left[\hat{\rho}\,\hat{\Pi}_T(\Delta)\right].
\end{equation}
The sample space must include the event $N$ of no detection occurring in the interval $[0, T]$. So we require that 
\begin{equation}
\hat{\Pi}_T((0,T])\leq\hat I,
\end{equation}
so that $\hat{\Pi}_{N} = \hat I - \hat{\Pi}_T((0,T])$ is a positive operator.

For the same instrument and preparation, an extension of the duration $T$
  must preserve the probabilities of earlier records:
\begin{equation}
\hat{\Pi}_{T'}(\Delta)=\hat{\Pi}_T(\Delta),
\qquad T'\geq T,\quad \Delta\subset(0,T]. \label{consist}
\end{equation}
Merely choosing when to stop recording
cannot alter the statistics of records already produced.

A natural defining property of a time observable is covariance:
a time translation of the system must induce the corresponding
translation of its recorded values.
\begin{equation}
\hat{\Pi}_{T+s}(\Delta+s)
=\hat U(s)\hat{\Pi}_T(\Delta)\hat U(s)^\dagger,
\qquad s\geq0. \label{covar}
\end{equation}
Here $\hat U(s)$ implements time translation. The restriction $s \geq 0$ implements the idea that covariance is meaningful only after the detector has been activated (at $t = 0$).

In algebraic quantum field theory \cite{Haag}, each open spacetime region
$O$ is assigned an algebra $\mathcal A(O)$ of bounded operators,
interpreted as the observables localized in that region.
These algebras respect inclusion: if $O_1\subset O_2$, then
$\mathcal A(O_1)\subset\mathcal A(O_2)$. Relativistic locality
is expressed by requiring that operators associated with
spacelike-separated regions commute. Smeared field operators
provide the standard connection with conventional QFT:
bounded functions of self-adjoint fields smeared with test
functions supported in $O$ belong to $\mathcal A(O)$.
In this framework, a POVM effect is algebraically localized
in $O$ if it belongs to $\mathcal A(O)$. Identifying such
membership with the localization of a measurement record
is a common operational interpretation. Operational measurement frameworks based on local system–probe couplings likewise yield induced system observables localized in local algebras \cite{FeVe}.

However, an adaptation of a theorem by Giannitrapani \cite{Gianni} shows that this interpretation is not available for time observables.

\begin{theorem}
Let $\{\hat{\Pi}_T\}_{T>0}$ be a family of detection-time POVMs
on $(0,T]$, with $\hat{\Pi}_T((0,T])\leq\hat I$, satisfying Eqs. (\ref{consist}) and (\ref{covar}).
If $\Omega$ is a normalized invariant vector,
$\hat U(s)\Omega=\Omega$, then
$$
 \hat{\Pi}_T(\Delta) \Omega = 0
$$
for every $T>0$ and every Borel set $\Delta\subset(0,T]$.
\end{theorem}
\begin{proof}
By consistency, we may write
$\hat{\Pi}(\Delta):=\hat{\Pi}_T(\Delta)$ independently of $T$,
provided $\Delta\subset(0,T]$.

Fix $\tau>0$ and set
$
c_\tau
:=\langle\Omega|\hat{\Pi}((0,\tau])|\Omega\rangle\geq 0$.
Covariance and invariance of $\Omega$ imply
$
\langle\Omega|
\hat{\Pi}((k\tau,(k+1)\tau])
|\Omega\rangle=c_\tau
$,
for every integer $k\geq0$. Hence, by additivity and normalization,
$$
Nc_\tau
=\sum_{k=0}^{N-1}
\langle\Omega|
\hat{\Pi}((k\tau,(k+1)\tau])
|\Omega\rangle
=\langle\Omega|\hat{\Pi}((0,N\tau])|\Omega\rangle
\leq1,
$$
for every positive integer $N$. It follows that $c_\tau=0$.

Then, for any $\Delta \subset (0, T]$,
$$
0 \leq \langle\Omega|\hat{\Pi}(\Delta)|\Omega\rangle \leq \langle\Omega|\hat{\Pi}((0,T])|\Omega\rangle = c_{T} = 0.
$$
Since $\langle\Omega|\hat{\Pi}(\Delta)|\Omega\rangle = ||\sqrt{\hat{\Pi}_T(\Delta)}\Omega||^2 = 0$, we obtain $\sqrt{\hat{\Pi}_T(\Delta)}\Omega = 0$, hence, $\hat{\Pi}_T(\Delta)\Omega = 0$.
\end{proof}
This result applies to a general quantum theory. We specialise to  a QFT on Minkowski spacetime with 
Hilbert space $\mathcal H$.  The usual Reeh--Schlieder
property applies\cite{Haag}: for every nonempty open spacetime region $O$, the
vacuum $\Omega$ is cyclic for $\mathcal A(O)$, i.e., $\overline{\mathcal A(O)\Omega}=\mathcal H$.
By locality, $\Omega$ is  {\em separating} for
$\mathcal A(O)$ whenever the spacelike complement of $O$
contains a nonempty open set: If 
$
\hat A\in\mathcal A(O)$  and $\hat A\Omega=0$
then $\hat A = 0$.
In particular, this holds for every bounded region $O$. 

Theorem 1 then implies: 

\begin{corollary}
Every nonzero detection-time
effect satisfying the assumptions of Theorem 1 lies outside
$\mathcal A(O)$ for every bounded region $O$.
\end{corollary}

Suppose that a detector is located in a spatial region $Q$. The corollary asserts that a measurement that identifies a record in the  spacetime region $O_{\Delta} = \Delta \times Q$, where $\Delta \subset [0, T]$ cannot be described by positive operators in  $\mathcal A(O_{\Delta})$. 

The assumptions of Theorem~1 can be relaxed. The obstruction
does not depend on continuous time, exact normalization,
or exact covariance. In particular, no covariance condition
is required for intervals shorter than an arbitrary scale $\tau_0$.
For each interval width $\tau>\tau_0$, covariance may hold
with a nonzero error, provided its norm, averaged over time
translations, vanishes in the long-time limit.
The precise conditions and proofs are given in the Appendix.

Existing constructions of detection-time observables illustrate how algebraic nonlocality arises and where the assumptions of Theorem 1 may fail.

\begin{enumerate}

\item Glauber's photodetection theory \cite{Glauber1} defines an  unnormalized measure  for detection in a spacetime region $D$. 
For a free scalar field,  
\begin{equation}
\hat{\Pi}(D) = \int_{D} d^4x \hat\phi^{(-)}(x)\hat\phi^{(+)}(x), \nonumber
\end{equation}
where $\hat\phi^{(+)}$ and $\hat\phi^{(-)}$ are the annihilation
and creation parts of the field, respectively. The positive operator $\hat{\Pi}(D)$ is generally unbounded on the full
Fock space and it therefore does not define either a POVM or an element of a local algebra, although
it can define a bounded operator on suitable energy-restricted
sectors.

Since $\hat\phi^{(+)}(x)\Omega=0$, the vacuum response vanishes. However, the fields
$\hat\phi^{(\pm)}(x)$ are non-local functionals of the fundamental field  $\hat\phi(x)$. Hence, $\hat{\Pi}(D)$ does not belong in any local algebra.

\item   Brunetti and Fredenhagen~\cite{Freden}
start from a positive
operator $\hat A$, and they define
\begin{equation}
\hat B(\Delta)=\int_\Delta dt\,\alpha_t(\hat A),
\qquad
\hat B=\int_{\mathbb R}dt\,\alpha_t(\hat A),
\end{equation}
where $\alpha_t$ stands for the time evolution automorphism. Then, they 
  obtain a covariant occurrence-time POVM through operator
normalization,
$\hat \Pi(\Delta)=\hat B^{-1/2}\hat B(\Delta)\hat B^{-1/2}$.

The construction excludes  
the sector with divergent $\hat B$. Such a divergence does
not by itself specify a measurement outcome, so the
operational meaning of the excluded states is unclear.
For QFT, the authors suggest using an almost-local operator $\hat{A}$.
The full-time normalization does not establish  algebraic
localization of the resulting effects. Replacing it by a
 normalization in the interval $[0, T]$ generally introduces a dependence
on $T$ and therefore violates the consistency
condition (\ref{consist}).

\item The Quantum Temporal Probabilities (QTP) framework \cite{QTP1, QTP2, QTP3} includes the apparatus in the quantum description, assuming local coupling to the measured field. 
Detection events correspond to decoherent, coarse-grained
apparatus records. The detection time variable is also coarse-grained and it is sharply distinguished from the time parameter of Heisenberg evolution.
 To leading order in the detector--field
coupling, the unnormalized detection effects take the form
\begin{equation}
\hat{\Pi}(D)
=\int d^4x\,d^4x'\,
R^{ab}(D;x,x')\,
\hat C_a(x)\hat C_b(x'),
\end{equation}
where $\hat C_a(x)$ are local composite field operators and
$R^{ab}$ encodes the detector response and self-dynamics as well as the coarse-graining of the record.  
The region $D$ labels the location of the recorded event.

The expression yields a finite  nonzero vacuum
false-alarm rate, whose accumulated contribution grows linearly
with recording duration and therefore fails the bound
(\ref{bound}). 
POVMs are constructed by removing the vacuum contribution, and they lead  to meaningful detection-time distributions
for relativistic particles of different species \cite{QTP1, QTP4}.   Unlike
Glauber's construction, QTP uses local fields but the kernel $R^{ab}$  
does not restrict the field arguments $x,x'$ to the recording
region $D$. The locality of the field--apparatus coupling
does not entail algebraic localization of the detection effect.
\end{enumerate}

Our result has important implications for QFT measurement theory. 
 If detection times admit a covariant POVM
description, including the approximate form considered in the Appendix,
then their nonzero effects cannot be algebraically local
observables. Avoiding this conclusion requires either abandoning
the POVM description of recorded detection times or explaining
 in what sense a POVM that isn't even approximately covariant can represent detection time.

Accepting algebraically nonlocal effects does not contradict
the Haag--Kastler axioms, which do not explicitly require every
physical measurement to correspond to a local observable. Indeed, particle detectors are often represented by 
approximately local observables, in order to
reconcile macroscopic localization with zero vacuum response
\cite{Haag}. 
Our result nevertheless challenges the identification of algebraic
localization with the physical localization of measurement
records, and raises the question of what information local
algebras provide about complete experimental outcomes.

The scope of this challenge extends beyond experiments dedicated to temporal observables.
All measurements produce macroscopic records---for
example, an electric pulse in a particle detector---whose emergence
can be correlated with the simultaneous reading of a local clock.
The resulting record includes a timestamp $t$ alongside the
other recorded variables $q$, and can be described by a joint
POVM, written schematically as $\hat{\Pi}(t,q)$.
If the timestamp is disregarded, the remaining statistics are
described by the marginal $\int dt\,\hat{\Pi}(t,q)$.
However, whenever the temporal marginal
$\int dq\,\hat{\Pi}(t,q)$ satisfies the assumptions of Theorem~1,
the joint POVM cannot take all its values in any bounded-region
local algebra. We are not aware of any measurement producing distinct individual
records that is exempt from this issue: each record can be
timestamped, bringing its complete description within the scope
of the obstruction.

This suggests distinguishing between the spacetime coordinates of local fields and those that appear in detection POVMs. The former specify algebraic localization in QFT; the latter specify the physical localization of recorded events. These are distinct concepts: neither implies the other. This mathematical and conceptual distinction is explicit in the QTP approach \cite{QTP3}, where their relation depends on the apparatus response, the coarse graining, and the decoherence conditions required for definite macroscopic records \cite{GeHa2}.

The notion of causality in QFT measurements is directly affected. If physical measurements corresponded to local algebras, then spacelike commutativity of local algebras---the microcausality axiom---would suffice to demonstrate that QFT respects the {\em non-signalling condition}\cite{Earman}: choosing which measurement to perform in a region $O_1$ does not affect the outcome of measurements in  spacelike separated region $O_2$. 

The standard non-signalling argument assumes that a nonselective
measurement in a region $O_1$ is represented by Kraus operators
$\hat K_\alpha\in\mathcal A(O_1)$, satisfying
$\sum_\alpha\hat K_\alpha^\dagger\hat K_\alpha=\hat I$.
For an effect $\hat B\in\mathcal A(O_2)$, where $O_2$ is
spacelike separated from $O_1$, microcausality gives
$\sum_\alpha\hat K_\alpha^\dagger\hat B\hat K_\alpha
=\hat B\sum_\alpha\hat K_\alpha^\dagger\hat K_\alpha=\hat B$.
Thus the operation leaves all outcome probabilities in $O_2$
unchanged. Obviously, this proof does not work for measurements outside local algebras.

 Non-signalling must instead
be formulated as a property of joint detection probabilities in setups that involve multiple detectors. Its analysis 
 therefore requires
joint POVMs
$\hat{\Pi}(D_1,\ldots, D_n)$ for multiple detection events, at different spacetime regions.
 QTP provides constructions of such
multi-event probabilities \cite{QTP3}, but establishing their causal
properties requires control of the approximations and 
coarse graining  entering their derivation \cite{AnPl1}.

Detection times thus provide a concrete test for a complete
measurement theory of QFT. Such a theory must explain how
locally produced, timestamped records are represented when
their effects are not algebraically local, and establish
non-signalling for the corresponding experimental probabilities.
The result leaves open how this should be achieved, but shows
that algebraic localization and microcausality are, by themselves, insufficient.

 \section*{Acknowledgements}
 K.S. acknowledges support from the Andreas Mentzelopoulos Foundation through the  Branded Academic Position on ``Quantum Science and Technology" at the University of Patras.
 Both authors acknowledge the COST Action CA23115 ``Relativistic Quantum Information".

\begin{appendix}
\section{Relaxing the assumptions of Theorem 1}

The conclusion of Theorem~1 persists under several relaxations
of its assumptions. Throughout this section, we  assume the conditions of Theorem 1, and use the definitions given in the proof.  

\paragraph{Sublinear bounds instead of normalization.}
The normalization bound may be replaced by
\begin{equation}
\hat{\Pi}_T((0,T])\leq C(T)\hat I,
\qquad
\lim_{T\to\infty}\frac{C(T)}{T}=0.
\label{bound}
\end{equation}
With this assumption, we obtain $Nc_\tau \leq C(N\tau)$, from which the condition $c_{\tau} = 0$, and the rest of the theorem follows.

\paragraph{Discrete time and finite resolution.}
Let $\tau>0$ be a fixed temporal resolution, and let
$\hat{\Pi}_n$ be the positive effect for detection in the
$n$th bin. Assume
\begin{equation}
\hat{\Pi}_n
=\hat U(n\tau)\hat{\Pi}_0\hat U(n\tau)^\dagger,
\qquad
\sum_{n=0}^{N-1}\hat{\Pi}_n\leq C_N\hat I,
\qquad
\frac{C_N}{N}\longrightarrow0.
\end{equation}
Since $\Omega$ is invariant,
\begin{equation}
N\langle\Omega|\hat{\Pi}_0|\Omega\rangle
=\sum_{n=0}^{N-1}
\langle\Omega|\hat{\Pi}_n|\Omega\rangle
\leq C_N. \nonumber
\end{equation}
Taking $N\to\infty$ gives
$\langle\Omega|\hat{\Pi}_0|\Omega\rangle=0$.
Covariance and positivity then imply
$\hat{\Pi}_n\Omega=0$ for every $n$.
The argument therefore requires neither continuous time
nor arbitrarily fine resolution, only that the 
number of possible recording bins have no fixed upper bound.

\paragraph{Covariance after an initial transient.}
Suppose covariance is required only for intervals lying
after a fixed time $t_0$. The proof of Theorem 1 passes unchanged for times $t > t_0$, if we substitute the intervals $(0, N\tau]$ with $(t_0, t_0 + N \tau]$.

\paragraph{Approximate covariance.}
Exact covariance can be replaced by a condition on its
average error. For every $\tau>0$, let
\begin{equation}
\epsilon_\tau(s)
=
\left\|
\hat U(s)\hat{\Pi}((0,\tau])\hat U(s)^\dagger
-\hat{\Pi}((s,s+\tau])
\right\|,
\end{equation}
and assume that there exists $\tau_0 >0$, such that for all $\tau > \tau_0$,
\begin{equation}
\lim_{T\to\infty}
\frac{1}{T}\int_0^T ds\,\epsilon_\tau(s)=0.
\end{equation}

Define $\mu(\Delta)
=\langle\Omega|\hat{\Pi}(\Delta)|\Omega\rangle \leq 1$ and  $c_\tau=\mu((0,\tau])$.
Vacuum invariance implies that $\left|c_\tau-\mu((s,s+\tau])\right|
\leq\epsilon_\tau(s)$.
Hence,
\begin{equation}
Tc_\tau
\leq
\int_0^T ds\,\mu((s,s+\tau])
+\int_0^T ds\,\epsilon_\tau(s). \nonumber
\end{equation}
Writing $\mu(t)dt = \int \mu((t,+dt])$, we find
\begin{align}
\int_0^T ds\,\mu((s,s+\tau]) =   \int_{(0,T+\tau]} dt \int_0^T ds \mu(t) \chi_{[s,s+\tau]}(t) \leq \tau \int_{(0,T+\tau]} dt \mu(t) = \tau\,\mu((0,T+\tau])
\leq\tau. \nonumber
\end{align}

Consequently, $0\leq c_\tau
\leq\frac{\tau}{T}
+\frac{1}{T}\int_0^T ds\,\epsilon_\tau(s)
\longrightarrow 0$, and the conclusion of Theorem 1 follows.

\end{appendix}

\end{document}